\documentclass[prl,twocolumn,showpacs,amsmath,amssymb,floatfix]{revtex4}

\usepackage{graphicx}
\usepackage{dcolumn} 
\usepackage{bm}      
\usepackage{color}

\begin{document}

\title{Energy spectra stemming from interactions of Alfv\'en waves and turbulent eddies}

\author{P.D. Mininni$^{1,2}$ and A. Pouquet$^{2}$}
\affiliation{$^1$ Departamento de F\'\i sica, Facultad de Ciencias Exactas y 
         Naturales, Universidad de Buenos Aires, Ciudad Universitaria, 1428 
         Buenos Aires, Argentina. \\
             $^2$ NCAR, P.O. Box 3000, Boulder, Colorado 80307-3000, U.S.A.}
\date{\today}

\begin{abstract}
We present a numerical analysis of an incompressible decaying magnetohydrodynamic turbulence run on a grid of $1536^3$ points. The Taylor Reynolds number at the maximum of dissipation is $\approx 1100$, and the initial condition is a superposition of large scale ABC flows and random noise at small scales, with no uniform magnetic field. The initial kinetic and magnetic energies are equal, with negligible correlation. The resulting energy spectrum is a combination of two components, each moderately resolved. Isotropy obtains in the large scales, with a spectral law compatible with the Iroshnikov-Kraichnan theory stemming from the weakening of nonlinear interactions due to Alfv\'en waves; scaling of structure functions confirms the non-Kolmogorovian nature of the flow in this range. At small scales, weak turbulence emerges with a $k_{\perp}^{-2}$ spectrum, the perpendicular direction referring to the local quasi-uniform magnetic field. 
\end{abstract}

\pacs{47.27.Jv, 47.65.-d, 95.30.Qd, 94.05.Lk}
\maketitle

Magnetic fields permeate the universe, and with the increased resolving capabilities of instruments, a flurry of details on highly complex flows emerge.
The origin of such magnetic fields, the dynamo problem, is still not fully understood, in particular when the magnetic Prandtl number $P_M$ -- the ratio of kinematic viscosity $\nu$ to magnetic diffusivity $\eta$ -- differs substantially from unity, as it does in the  interstellar medium ($P_M\gg 1$) or in convective regions of stars and liquid cores of planets ($P_M\ll 1$). Numerous theoretical and numerical studies have been written (see \cite{Brandenburg05}), and recent laboratory experiments address this fundamental problem as well.
Here, we take a different approach by assuming that a dynamo mechanism works, presumably leading to approximate equipartition of the magnetic and kinetic energies, as observed for example in the solar wind. We then ask questions about the nature of the nonlinear dynamics of such a flow. The complete problem (a dynamo taken all the way to the nonlinear stage at a high Reynolds number as encountered in nature, and for $P_M$ differing substantially from unity) is still out of reach, but it might be among the key outcomes of petascale computing facilities and experimental devices currently being developed.
We specifically address in this paper the nature of the energy spectra that arise in such conditions. A more detailed account of the development and evolution (up to the 
peak of enstrophy) of structures and correlations of the flow will be presented elsewhere.

The magnetohydrodynamic (MHD) equations read:
\begin{eqnarray}
&& \frac{\partial {\bf v}}{\partial t} + {\bf v} \cdot \nabla {\bf v} = 
    -\frac{1}{\rho_0} \nabla {\cal P} + {\bf j} \times {\bf b} + \nu \nabla^2 
    {\bf v} , 
\label{eq:MHDv} \\
&& \frac{\partial {\bf b}}{\partial t} = \nabla \times ( {\bf v} \times
    {\bf b}) +\eta \nabla^2 {\bf b} ;
\label{eq:MHDb} \end{eqnarray}
${\bf v}$ is the velocity, ${\bf b}$ is the magnetic field, ${\bf j} = \nabla \times {\bf b}$ is the current density, ${\cal P}$ is the pressure, $\rho_0=1$ the density, $\nu=\eta=2 \times 10^{-4}$, and ${\bf \nabla} \cdot {\bf v} = \nabla \cdot {\bf b} = 0$.
One can also write these equations in terms of the Els\"asser variables ${\bf z}^\pm = {\bf v} \pm {\bf b}$, of energy $E^{\pm}$ and flux $\epsilon^{\pm}$.
We solve Eqs. (\ref{eq:MHDv}, \ref{eq:MHDb}) in a three-dimensional box of length $L_0=2\pi$ using periodic boundary conditions and a pseudospectral method, dealiased by the standard 2/3 rule; minimum and maximum wavenumbers are $k_{min}=1$ and $k_{max}=N^{1/3}/3$, with $N=1536^3$ grid points. At all times $k_D/k_{max}<1$, where $k_D$ is the dissipation wavenumber. The initial conditions for the velocity and magnetic fields are constructed from a superposition of three Beltrami (helical) flows to which smaller-scale random fluctuations are added with initial kinetic and magnetic energy $E_V = E_M = 0.5$ (with respective spectra $E_V(k)$ and $E_M(k)$); magnetic helicity $H_M = \left< {\bf a} \cdot {\bf b} \right> \approx 0.45$ (${\bf b} = \nabla \times {\bf a}$ where ${\bf a}$ is the vector potential, and the brackets denote volume average), and $\cos({\bf v},{\bf b}) = \left<{\bf v} \cdot {\bf b} \right> \left<|{\bf v}||{\bf b}| \right>^{-1} \approx 10^{-4}$ (see \cite{Mininni06} for details). The computation is stopped when the growth of the total dissipation saturates ($t=3.7$), at which time the Reynolds number based on the mechanical integral scale is $R_e = UL_V/\nu \approx 9200$, and that based on the mechanical Taylor scale is $R_{\lambda} = U\lambda_V /\nu \approx 1100$; $U$ is the r.m.s. velocity, the integral scales are defined as $L_i = 2\pi E_i^{-1} \int k^{-1} E_i(k)dk $, the Taylor scales are $\lambda_i = 2\pi ( E_i / \int k^2 E_i(k)dk )^{1/2} $, and $i$ is either $V$ or $M$. These scales at $t=3.7$ are $L_V \approx 2.6$, $\lambda_V \approx 0.31$, $L_M \approx 3.1$, and $\lambda_M \approx 0.39$. It was shown in \cite{Mininni06} that at early times, current and vorticity sheets form, and further roll-up, fold, or pile-up. Here we focus on the fully 
developed regime close to the peak of enstrophy. 

\begin{figure}
\includegraphics[width=9cm]{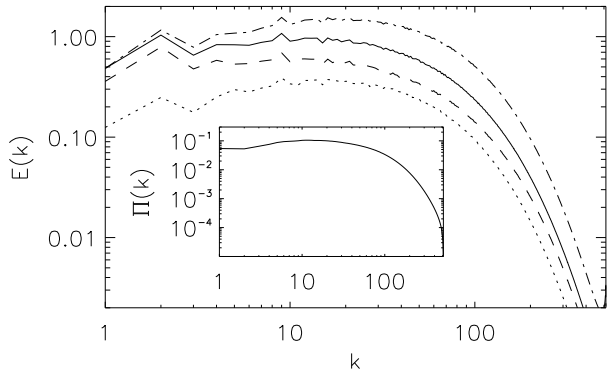}
\caption{
Energy spectra $E(k)$ (solid), $E_M(k)$ (dash) and $E_V(k)$ (dot) compensated by $k^{-3/2}$;
 $E(k)$ compensated by $k^{-5/3}$ is also shown (dash-dot).
 In the inset is the total energy flux $\Pi(k)$, indicating the extent of the inertial range.}
\label{fig:fluxl} \end{figure}

The total energy flux $\Pi(k)$ as a function of wavenumber at $t=3.7$ is shown in Fig. \ref{fig:fluxl} (inset), together with the total, kinetic, and magnetic energy spectra ($E(k)=E_V(k)+E_M(k)$), compensated by either $k^{-5/3}$  (Kolmogorov spectrum, hereafter K41) or $k^{-3/2}$ (Iroshnikov-Kraichnan, or IK \cite{Iroshnikov63}); the latter takes into account the slowing down of nonlinear transfer because of Alfv\'en waves. The scaling of $E(k)$ and $E_M(k)$ is close to 
$k^{-3/2}$ in a range of scales covering almost a decade. Energy spectra in MHD have been reported in the literature, both for decaying and forced flows, for data stemming from either numerical simulations or observations (mostly in the solar wind). In most cases, a spectrum close to K41 is found, although there are times when the IK solution is preferred \cite{Muller05}, and in some cases both scalings may be observed \cite{Chapman06}. Moreover, it is not clear whether universality obtains (in which case, for very large Reynolds numbers it would either be K41 or IK), or whether there are classes of universality, the boundaries to which are not necessarily known (see, e.g., the discussion in \cite{Kaneda07ed}). Note that different spectra have already been observed for different forcings in reduced MHD (RMHD, an approximation valid in the presence of a strong magnetic field) \cite{Dmitruk03}, and recently also reported for MHD runs with a strong imposed magnetic field. Note that the flow in our run has the largest $R_{\lambda}$ obtained in a direct numerical simulation, at the expense, though, of not being stationary.

\begin{figure}
\includegraphics[width=9cm]{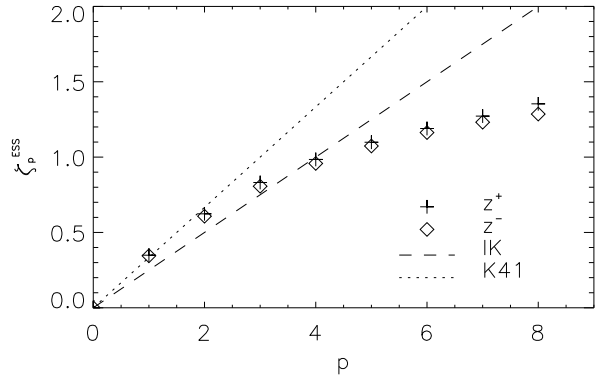} 
\caption{Scaling exponents $\zeta_p^\pm$ for ${\bf z}^\pm={\bf v}\pm{\bf b}$. The K41 (dot) and IK (dash) predictions are also shown. Note $\zeta^\pm_4\sim 1$.}
\label{fig:zetap} \end{figure}

To clarify the issue of which spectra should arise in MHD turbulence, at least two things can be done. On one hand, one computes isotropic high-order structure functions, in which case the differentiation between K41 and IK is enhanced, as already found in \cite{Politano98}. On the other hand, one can quantify the degree of anisotropy in the flow, something that will be discussed later. 
We define the longitudinal (parallel to the displacement ${\bf l}$) structure function of order $p$ for the quantity ${\bf f}$ as
$S_p^{\bf f}(\ell) = \left< \left[ \delta f_L ({\bf l}) \right]^p \right> = \left< \left\{ \left[  {\bf f}({\bf x}) - {\bf f}({\bf x}+{\bf l}) \right] \cdot {\bf l}/\ell \right\}^p \right>$.
If the field is self-similar we expect a scaling $S_p^{\bf f}(\ell) \sim \ell^{\zeta_p^{\bf f}}$  in the inertial range, where $\zeta_p^{\bf f}$ are the scaling exponents. In hydrodynamic turbulence, K41 theory predicts $\zeta_p=p/3$, and in MHD the IK theory leads to $\zeta_p=p/4$. In practice, these relations are modified by intermittency corrections due to extreme events at the small scales. The exact scaling laws for MHD turbulence derived in \cite{Politano98a}, $\left< \delta z_L^\mp ({\bf l}) | \delta {\bf z}^\pm ({\bf l}) |^2 \right> = -4 \epsilon^\pm \ell/3 $, can be used to define the inertial range and to improve the estimate of the scaling exponents, as is often done in hydrodynamics when the extended self-similarity (ESS) hypothesis is used \cite{Benzi93a}. These scaling laws, and structure functions up to order 8, were computed at $t=3.7$ for increments in the $\hat{x}$, $\hat{y}$, and $\hat{z}$ directions, and averaged. Figure \ref{fig:zetap} shows the resulting scaling exponents $\zeta_p^\pm$ for the Els\"asser fields ${\bf z}^\pm$, computed in the range in which the exact scaling laws hold.

Consistent with the computation of the energy spectra, the scaling of ${\bf z}^\pm$ is closer to the IK prediction, with $\zeta_3^+ = 0.831 \pm 0.006$, $\zeta_3^- = 0.806 \pm 0.005$, $\zeta_4^+ = 0.985 \pm 0.008$, $\zeta_4^- = 0.958 \pm 0.007$. As stated in \cite{Politano98a}, the exact scaling laws in MHD involve correlations between the velocity and magnetic fields (or equivalently between ${\bf z}^{\pm}$), and such correlations play a role in the dynamics of the flow. In particular, they can induce a shift from a standard K41 coupling; this effect has been modeled in \cite{Boldyrev06} as a scale variation of the angle between the velocity and magnetic field, i.e. in the relative cross helicity $\cos({\bf v},{\bf b})$. From the values of $\zeta_3^\pm$ and $\zeta_4^\pm$, the possibility of a Kolmogorov scaling hidden by a bottleneck (as reported in \cite{Haugen04}) can be ruled out for this simulation. However, even when the scaling exponents in the inertial range for $p = 3$ and 4 are closer to the IK prediction (with strong corrections due to intermittency), it is worth mentioning that the structure functions after using ESS do not show the same scaling at all scales. This is different from the hydrodynamic case where, after using the ESS hypothesis, a unique scaling at all scales is observed, even in the dissipative range. This brings us to our second point, linked to anisotropy.

Indeed, it is known that in MHD anisotropy develops at small scales under the influence of a large-scale magnetic field. Several models have been written to explain how this might affect the small-scale dynamics. The weak turbulence (WT) theory \cite{Galtier00} leads to an exact spectrum $\sim k_{\perp}^{-2}f(k_\parallel)$, where $k_{\parallel}$ and $k_{\perp}$ are the amplitudes of the wave vectors in the directions parallel and perpendicular to a uniform magnetic field ${\bf B}_0$. It has long been argued that at high Reynolds number, leading to a large scale separation, the large-scale magnetic field ${\bf B}_\textrm{LS}$ plays the same dynamical role as ${\bf B}_0$. As a result, we can expect the small scales to be anisotropic with respect to ${\bf B}_\textrm{LS}$. However, since ${\bf B}_\textrm{LS}$ evolves in time with a characteristic correlation time $\tau_\textrm{LS}$ and has a curvature proportional to the Taylor scale $\lambda_M$, in the absence of a uniform field and for isotropic initial conditions, one may assume that isotropy is still preserved in the large and intermediate scales ($\ell \gtrsim \lambda_M$).

\begin{figure}
\includegraphics[width=9cm]{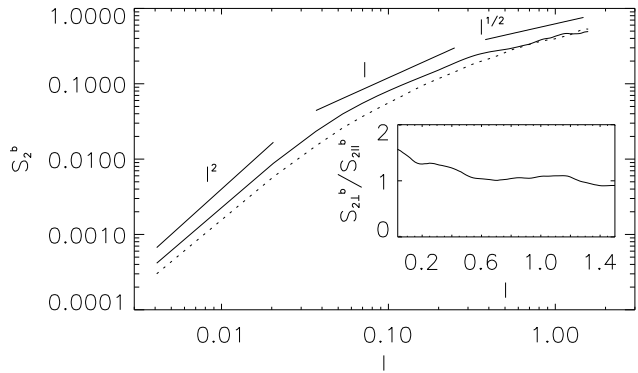}
\caption{Second order structure functions $S_{2\perp}^{\bf b}(\ell)$ 
    (solid) and $S_{2\parallel}^{\bf b}(\ell)$ (dot). Several slopes are 
    given as a reference. The inset shows the ratio 
    $S_{2\perp}^{\bf b}(\ell)/S_{2\parallel}^{\bf b}(\ell)$ in linear 
    coordinates.}
\label{fig:angle} \end{figure}

To illustrate this, we follow the procedure given in \cite{Milano01}. A local mean field ${\bf B}_\textrm{LS}({\bf x})$ is computed as the average of the magnetic field ${\bf b}$ in a box of size $L_M$ with center on ${\bf x}$, such as 
${\bf B}_\textrm{LS} = \left< {\bf b} \right>_{\textrm{box}(L_M,{\bf x})}$.
Then, unit vectors ${\bf l}_\parallel$ and ${\bf l}_\perp$ ($\parallel$ and $\perp$ with respect to ${\bf B}_\textrm{LS}$) are defined, and the two following longitudinal magnetic structure functions
\begin{equation}
S_{2\perp}^{\bf b}(\ell) = S_2^{\bf b}(\ell \, {\bf l}_\perp) ,  \; \; \; 
S_{2\parallel}^{\bf b}(\ell) = S_2^{\bf b}(\ell \, {\bf l}_\parallel) ,
\end{equation}
are computed for all points in the vicinity of ${\bf x}$ and for $\ell \in [2 \pi k_{max}^{-1},L_M/2]$. The process is repeated for several values of ${\bf x}$, and $S_{2\perp}^{\bf b}$ and $S_{2\parallel}^{\bf b}$ are averaged over all the boxes (the results shown are the average over $\approx 10^7$ points). Figure \ref{fig:angle} gives the resulting $S_{2\perp}^{\bf b}$ and $S_{2\parallel}^{\bf b}$, as well as their ratio. In the limit $\ell \to 0$, this ratio can be associated with one of the so-called Shebalin angles \cite{Shebalin83}
\begin{equation}
\tan ^2(\theta) = 2 \lim_{\ell \to 0} \frac{S_{2\perp}^{\bf b}(\ell)}
    {S_{2\parallel}^{\bf b}(\ell)} ,
\label{eq:shebalin} \end{equation}
which measures the anisotropy of ${\bf b}$ with respect to the local mean field.

\begin{figure}
\includegraphics[width=9cm]{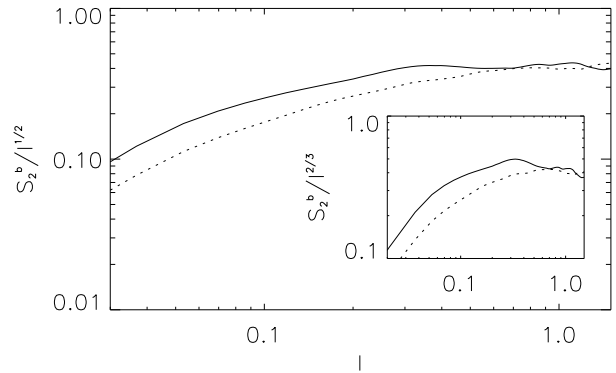}
\caption{$S_{2\perp}^{\bf b}$ (solid) and $S_{2\parallel}^{\bf b}$ (dot) compensated by $\ell^{1/2}$. The inset shows the same quantities compensated by $\ell^{2/3}$.}
\label{fig:aniWT} \end{figure}

We see that scales smaller than $\ell \approx 0.4$ are anisotropic, in agreement with \cite{Milano01}; $S_{2\perp}^{\bf b}(\ell)> S_{2\parallel}^{\bf b}(\ell)$ in this range. However, isotropy obtains in an intermediate range of scales corresponding roughly to the $-3/2$ range displayed in Fig. \ref{fig:fluxl}. For very small increments ($\ell \leq 0.02 \approx 2\pi k_D^{-1}$), 
both structure functions follow $\sim \ell^2$ scaling, as expected for smooth fields in the dissipative range. At scales larger than the dissipation scale, but smaller than $\ell \approx 0.4$, a range with $S_{2\perp}^{\bf b} \sim \ell$ is observed, consistent with a $k_\perp^{-2}$ scaling in the energy spectrum as predicted by WT theory \cite{Galtier00}. In the range $\ell \approx [0.4,1.2]$, where fluctuations are roughly isotropic, $S_{2\perp}^{\bf b} \sim S_{2\parallel}^{\bf b} \sim \ell^{1/2}$. Figure \ref{fig:aniWT} shows the structure functions compensated by $\ell^{1/2}$ (for IK scaling) and $\ell^{2/3}$ (for K41). The $\sim \ell^{1/2}$ scaling in this range is again consistent with the isotropic energy spectrum and scaling of the isotropic ($p=3$ and 4) structure functions discussed previously.

The recovery of isotropy observed at intermediate scales is not obvious. Note that $\tan^2(\theta) > 1$, and for very small $\ell$, 
we have $S^{\bf b}_{2\perp} \sim S^{\bf b}_{2\parallel}\sim \ell^2$. If ${\bf B}_\textrm{LS}$ is strong, and there is enough scale separation for the local mean field to look as a uniform field for the small scales, WT theory is consistent with $S_{2\parallel}(\ell) \sim \ell^{1/2}$. Now assume $S_{2\perp}(\ell) \sim \ell^\alpha$. For $\alpha \ge 1/2$, the structure functions satisfy $S_{2\perp}/S_{2\parallel}\ge 1$ at all scales. Thus, the recovery of isotropy at intermediate scales should not be associated with the small scale fluctuations but rather with the fluctuations of the large scale magnetic field, and thus may not always take place. This is consistent with results from MHD with a dc magnetic field: in this case, the fields are anisotropic at all scales \cite{Milano01}. It may mean that in MHD the energy spectrum is not universal and depends on some property of the large scale field \cite{Kaneda07ed}. Extensive high resolution parametric studies are needed to determine conclusively what is happening.

Let $\lambda_M$ be the largest scale for the fluctuations to see the mean magnetic field as a uniform field; for $\ell \approx \lambda_M$, the curvature of the large scale magnetic field cannot be neglected, and fluctuations become more isotropic. At scales larger than $\lambda_M$, we thus have isotropic fluctuations with a local mean magnetic field ${\bf B}_\textrm{LS}$. Assuming equipartition and constant energy flux $\varepsilon \sim u_\ell^2 \tau_{\textrm{A},\ell} / \tau_{\textrm{NL},\ell}^2$ (where $\tau_{\textrm{NL},\ell} \sim \ell/v_\ell$ is the nonlinear turnover time, and $\tau_{\textrm{A},\ell} \sim \ell/B_\textrm{LS}$ is the Alfv\'en crossing time), we obtain $b_\ell \sim v_\ell \sim \ell^{1/2}$ and the IK energy spectrum $E(k) \sim k^{-3/2}$. At scales smaller than $\lambda_M$, the fluctuations are anisotropic and several phenomenologies have been put forward \cite{Goldreich95,Boldyrev06,Galtier00}. Assuming $v_l$ and $b_l$ are mostly in the direction perpendicular to ${\bf B}_\textrm{LS}$, $\tau_{\textrm{NL},l} \sim \ell_\perp/v_l$. The anisotropic extension of IK takes the Alfv\'en time as $\tau_{\textrm{A},l} \sim \ell_{\parallel}/B_\textrm{LS}$, which results in $v_l \sim b_l \sim \ell_\perp \ell_\parallel^{-1/2}$ and the energy spectrum $E(k) \sim k_{\perp}^{-2}k_{\parallel}^{1/2}$. In the context of this run, where the correlation length of the large scale magnetic field is $L_M$, we can assume $\tau_{\textrm{A},l} \sim L_M/B_\textrm{LS}$, which leads to the same $\sim \ell_\perp$ and $\sim k_\perp^{-2}$ scaling. Note that this $\perp$ spectrum is what WT theory predicts, and that for the isotropic case ($k_{\perp}\sim k_{\parallel}$), $E(k) \sim k_{\perp}^{-2}k_{\parallel}^{1/2}$ turns into the IK spectrum. In this light, it is not surprising that the IK spectrum at large scales be followed, in the same dynamical vein, by its wave turbulence anisotropic counterpart at small scales.

Another spectrum for MHD turbulence has been advocated in \cite{Goldreich95}, whereby the anisotropy of the flow induces a Kolmogorov spectrum. In the anisotropic range, this implies $S_{2 \perp}^{\bf b} \sim \ell^{2/3}$. Another anisotropic phenomenological model based on field alignment \cite{Boldyrev06} implies $S_{2 \perp}^{\bf b} \sim \ell^{1/2}$. Such scalings are not observed in our computation, and the results in Fig. \ref{fig:aniWT} suggest $S_{2 \perp}^{\bf b} \sim \ell$ is a better fit in the anisotropic range. However, other scaling laws cannot be ruled out completely for the following reason. It is well known that the assumption leading to WT (namely, that the characteristic time of the wave be shorter than the time for nonlinear exchange of energy between eddies) is non-uniform in scale, and breaks down at small scales when the turn-over time has become sufficiently small. It is possible that we have not achieved yet sufficient scale separation for this to occur, and simulations at higher Reynolds number are needed in order to reach a better and more detailed understanding of such flows.

Since we also know that MHD turbulence displays non-local transfer of energy \cite{Alexakis05a}, leading to a direct connection between the large and the small scales, a break-down of universality can be a candidate to explain the variety of solutions reported in the literature. The parameters that can influence such a break-down are many and will require further study. In this context, it is useful to compare the present results with simulations of forced MHD turbulence. In our simulations we can associate the scale where the transition from the isotropic to anisotropic field takes place with $\lambda_M$, and the correlation time of the local mean field $\tau_\textrm{LS}$ with $L_M/B_\textrm{LS}$. In a forced simulation, the correlation time of the large scale magnetic field can be associated with the correlation of the external forcing $\tau_\textrm{F}$. If $\tau_\textrm{F} \ll \tau_{\textrm{A},\ell}$, then the local mean field changes faster than the Alfv\'en wave crossing time at the scale $\ell$, and we should not expect anisotropy to develop at that scale, or wave interactions to be relevant. The scale where the transition takes place would change accordingly. Such a dependence would be consistent with results obtained in forced RMHD simulations \cite{Dmitruk03} and recent MHD simulations \cite{Mason07}.

\begin{acknowledgments}
We acknowledge discussions with D.C. Montgomery; PDM acknowledges discussions with D.O. G\'omez. The NSF-CMG grant 0327888 was instrumental in our work. Computer time was provided by the NCAR BTS program. PDM is a member of the Carrera del Investigador Cient\'\i fico of CONICET.
\end{acknowledgments}


\end{document}